\documentclass[aps,prb,superscriptaddress,showpacs,twocolumn]{revtex4-1}
\usepackage{graphicx}

\begin{document}

\title{Theoretical Study of Carbon Clusters in Silicon Carbide Nanowires}
\author{J. M. Morbec}
\email{jmmorbec@gmail.com}
\affiliation{Instituto de Ci\^encias Exatas, Universidade Federal de Alfenas, CEP 37130-000,
Alfenas, MG, Brazil}
\author{R. H. Miwa}
\affiliation{Instituto de F\'isica, Universidade Federal de Uberl\^andia, Caixa Postal 593, CEP
38400-902, Uberl\^andia, MG, Brazil}

\date{\today}

\begin{abstract}
  Using first-principles methods we performed a theoretical study of
  carbon clusters in silicon carbide~(SiC) nanowires. We examined small
  clusters with carbon interstitials and antisites in
  hydrogen-passivated SiC nanowires growth along the [100] and [111]
  directions. The formation energies of these clusters were calculated
  as a function of the carbon concentration. We verified that the
  energetic stability of the carbon defects in SiC nanowires depends
  strongly on the composition of the nanowire surface: the
  energetically most favorable configuration in carbon-coated [100]
  SiC nanowire is not expected to occur in silicon-coated [100] SiC
  nanowire. The binding energies of some aggregates were also
  obtained, and they indicate that the formation of carbon clusters in
  SiC nanowires is energetically favored.
\end{abstract}



\maketitle

\section{Introduction} \label{intro}

Silicon carbide (SiC) is a wide-band-gap semiconductor with excellent physical, electronic and mechanical
properties\cite{choyke} such as high thermal conductivity, high breakdown
field, low density, high saturation velocity, high mechanical strength, and stability at high
temperature. These exceptional features make SiC a promising candidate to replace silicon in
electronic devices operating in  
high-power, high-frequency, and high-temperature 
regimes.\cite{madar}

In the last years, SiC nanostructures (like nanospheres,\cite{shen} nanosprings,\cite{dzhang} 
nanowires\cite{ho} and nanotubes\cite{sun}) have been successfully synthesized, and several
theoretical and experimental works \cite{shen, dzhang, ho, sun, rurali, yan, melinon, shim, lsun}
have been performed to investigate their structural and electronic properties. The unique features of SiC
combined with quantum-size effects make the SiC nanostructures interesting materials for
nanotechnology applications. For instance, SiC nanowires and nanotubes have been considered as candidates for hydrogen
storage nanodevices\cite{melinon} and for building blocks in molecular electronic applications.\cite{appell} 
In particular, silicon carbide nanowires (SiC NWs) have excellent field emission properties,\cite{pan} 
high mechanical stability and high electrical conductance,\cite{rurali} and could be used as
nanoscale field emitters or nanocontacts in harsh environments.

Some optical and electronic properties of semiconductors may be modified by the presence of
defects. The most common defects in SiC are vacancies, interstitials, antisites and
clusters. These defects are mainly formed during the growth process and ion
implantation of dopants. 
Vacancies and interstitials of C and Si in 3C-, 4H- and 6H-SiC bulks have been thoroughly
investigated in theoretical and experimental works.\cite{zyw, bock, torpo, torpo1, torpo2, torpo3} These investigations have showed
that the C and Si vacancies are electron and hole traps,\cite{torpo, torpo1, torpo2, torpo3} and that C and Si interstitials have
higher mobility than vacancies, although the mobility of point defects in SiC is reduced as compared
to another semiconductors (like
silicon).\cite{zyw} The high mobility of carbon interstitials favors the formation of
carbon-interstitial clusters. Using {\it ab initio} methods, Gali et al.\cite{gali}
systematically investigated small clusters of carbon interstitials and antisites in 3C- and 4H-SiC 
bulks, and verified that the formation of carbon aggregates is energetically favored. 

In spite of some theoretical studies on carbon aggregates in SiC bulk,\cite{gali, matta, matta1} the investigation of carbon
clusters in SiC NWs is very scarce. Motivated by the lack of studies on C aggregates in SiC NWs and
by their energetically favorable formation in SiC bulk,  in this work we performed an {\it ab initio} study of
small carbon clusters in SiC NWs. We considered hydrogen-passivated SiC NWs grown along the [111]
and [100] directions and examined clusters with interstitial and antisites carbon atoms. 
The formation  energies of these clusters were determined as a function of the C concentration.  
We calculated the binding energies of some aggregates, and our results 
indicate that the formation of carbon clusters in SiC nanowires is energetically favored. 
Besides [111] and [100] SiC NWs, the carbon clusters
were also investigated in 3C-SiC bulk, in order to compare the effect of C defects in SiC bulk and
NW.

\section{Methodology} \label{method}

In this work we present {\it ab initio} calculations based on density functional theory\cite{dft}
(DFT) carried out by using the SIESTA code.\cite{siesta} We used local spin-density
approximation\cite{lsda, lsda1, ca} (LSDA) for the exchange-correlation functional and 
norm-conserving fully-separable pseudopotentials\cite{pseudo} to treat the electron-ion interactions. 
The Kohn-Sham orbitals were expanded using a linear combination of numerical pseudoatomic
orbitals\cite{lcao} and a double-zeta basis set with polarization functions\cite{dzp} (DZP) was employed 
to describe the valence electrons. 

The 3C-SiC bulk, and the [111] and [100] SiC NWs, were modeled within
the supercell approach, with 128, 232 and 279 atoms, respectively. The
SiC NWs were constructed from the 3C-SiC structure and the dangling
bonds of their surfaces were saturated with hydrogen atoms. Due to the
periodic boundary conditions, a vacuum region of about 10 \AA \ was
used to avoid interactions between a NW and its image. We have
considered axial lengths (along the NW growth direction) of about 15.2
and 13.2 \AA \ for the [111] and [100] SiC NWs, respectively.  The
geometries were optimized using the conjugated gradient scheme, within
a force convergence criterion of 0.05~eV$/$\AA.  The Brillouin zone
was sampled by using 2 special {\bf k} points for 3C-SiC bulk, and 1
special {\bf k} point for [100] and [111] SiC
NWs. 
 We verified the convergence of our total-energy results with respect to the number of special {\bf k}
points using up to 9 {\bf k} points for 3C-SiC bulk and 4 {\bf k} points for [100] and [111]  
SiC NWs.

\section{Results and discussion} \label{results}

\begin{figure}[!h]
\includegraphics[scale=0.6]{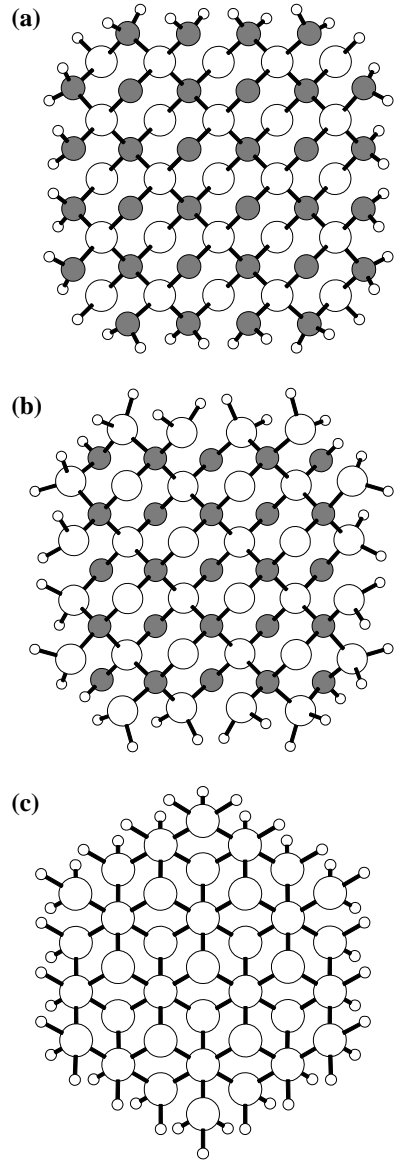}
\caption{\label{fig1}
Cross-section view of the structural models of 
(a) C-coated [100] SiC NW, (b) Si-coated [100] SiC NW and (c) [111] SiC NW. 
Carbon, silicon and hydrogen atoms are represented by filled, empty and small-empty circles, respectively.}
\end{figure}

We examined C clusters in 3C-SiC bulk and in hydrogen-passivated SiC
NWs grown along the [111] and [100] directions.\cite{rurali, rqzhang}  
We considered SiC NWs with diameter of about 10 \AA, constructed from the 3C-SiC
structure.\cite{zhang, ho}  For [100] SiC NWs, two kinds of wires were
studied: carbon-coated [100] SiC NW and silicon-coated [100] SiC NW,
whose surfaces are, respectively, carbon and silicon
terminated. Figure \ref{fig1} presents the cross-section view of the
structural models of carbon-coated [100] SiC NW [Fig.~\ref{fig1}(a)],
silicon-coated [100] SiC NW [Fig.~\ref{fig1}(b)] and [111] SiC NW
[Fig.~\ref{fig1}(c)]. Note that the [111] SiC NW surface has the same
number of Si and C atoms.

\begin{figure}[!h]
\includegraphics[scale=0.50]{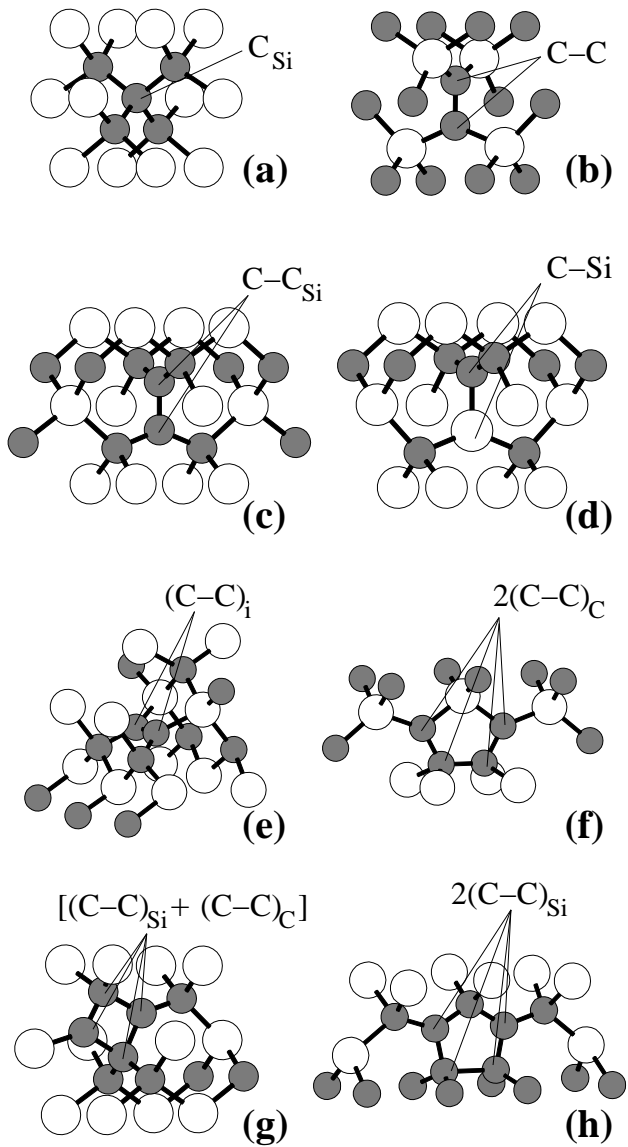}
\caption{\label{fig2}
Optimized geometries of the carbon defects investigated in 3C-SiC bulk and in 
hydrogen-passivated [100] and [111] SiC NWs. (a) C$_{\rm Si}$, (b) C -- C, (c) C -- C$_{\rm Si}$, (d) C -- Si, (e) (C -- C)$_{\rm i}$,
(f) 2(C -- C)$_{\rm C}$, (g) [(C -- C)$_{\rm Si}$ + (C -- C)$_{\rm C}$], and (h) 2(C -- C)$_{\rm
Si}$.}
\end{figure}

Eight C defects were investigated\cite{note1} in the present work (see
Fig. \ref{fig2}): C$_{\rm Si}$ (carbon antisite), C -- C
($\langle100\rangle$ split of C interstitial + C site), C -- C$_{\rm
  Si}$ ($\langle100\rangle$ split of C interstitial + C antisite), C
-- Si ($\langle100\rangle$ split of C interstitial + Si site), (C --
C)$_{\rm i}$, 2(C -- C)$_{\rm C}$, [(C -- C)$_{\rm Si}$ + (C --
C)$_{\rm C}$], and 2(C -- C)$_{\rm Si}$.

In order to examine the energetic stability of the C defects in the
3C-SiC bulk and in the [100] and [111] SiC NWs, we used the grand
canonical potential at $T=0$~K (as suggested in the
Refs.~\onlinecite{chadi} and \onlinecite{froyen})

\begin{equation}
\Omega=E_{tot}-\sum_i N_i\mu_i.
\label{eq1}
\end{equation}
Here $E_{tot}$ is the total energy of the considered structure,
$\mu_i$ is the chemical potential of atomic specie $i$ ($i=$~Si, C or
H) and $N_i$ is the number of atoms $i$ in the system.  In the
thermodynamic equilibrium,\cite{sabisch}
\begin{equation}
\mu_{\rm Si}+\mu_{\rm C}=\mu_{\rm SiC}^{\rm bulk}
\label{eq2}
\end{equation}
and
\begin{equation}
\mu_{\rm SiC}^{\rm bulk}=\mu_{\rm Si}^{\rm bulk}+\mu_{\rm C}^{\rm bulk}-\Delta H({\rm SiC}),
\label{eq3}
\end{equation}

\noindent
where $\Delta H({\rm SiC})$ is the formation heat of the SiC bulk.
Employing diamond structure\cite{grossner} for bulk phases of
Si and C, and zincblend structure\cite{grossner, kackell} for SiC, we
found $\Delta H({\rm SiC})=0.73$~eV, which is in good agreement with
the experimental value $\Delta H({\rm SiC})=0.68$~eV.\cite{harr} 

The upper limits of $\mu_{\rm Si}$ and $\mu_{\rm C}$ are $\mu_{\rm
  Si}^{\rm bulk}$ and $\mu_{\rm C}^{\rm bulk}$, respectively. Hence,
the potential fluctuations $\Delta \mu_{\rm Si}= \mu_{\rm Si} -
\mu_{\rm Si}^{\rm bulk}$ and $\Delta \mu_{\rm C}= \mu_{\rm C} -
\mu_{\rm C}^{\rm bulk}$ are restricted to
\begin{equation}
-\Delta H({\rm SiC}) \leq \Delta \mu_{\rm Si,\, C} \leq 0.
\label{eq4}
\end{equation}
This range defines the C-rich (or Si-poor, where $\Delta \mu_{\rm C} = 0$ and $\Delta \mu_{\rm Si} = -\Delta H{\rm
(SiC)}$) and Si-rich (or C-poor, where $\Delta \mu_{\rm C} = -\Delta H{\rm (SiC)}$ and $\Delta
\mu_{\rm Si} = 0$) limits. 

In Table \ref{tab1} we present the formation energies, relative to the
pristine structures, of C defects in 3C-SiC bulk, and in
hydrogen-passivated [100] and [111] SiC NWs. The two values presented
in each column correspond to Si-rich (or C-poor) and C-rich
limits. For 3C-SiC bulk at stoichiometric condition we found that the
formation energies of C$_{\rm Si}$, C -- C, C -- C$_{\rm Si}$, (C --
C)$_{\rm i}$, 2(C -- C)$_{\rm Si}$ and $[$(C -- C)$_{\rm Si}$ + (C --
C)$_{\rm C}$$]$ are 3.49, 7.38, 6.87, 8.79, 9.51 and 11.55~eV,
respectively. These results are in good agreement with those obtained
by Gali et al. in Ref.~\onlinecite{gali}.

\begin{table*}
  \caption{Relative formation energies ($\Delta \Omega$) of carbon defects 
    in 3C-SiC  bulk, and hydrogen-passivated [100] and [111] SiC NWs. In 
each column the two values correspond to  the Si-rich (or C-poor, 
where $\Delta \mu_{\rm C} = -\Delta H{\rm (SiC)}$) and C-rich 
($\Delta\mu_{\rm C} = 0$) limits (Si-rich/C-rich). The energies 
are in eV and we used $\Delta H{\rm (SiC)}=0.73$~eV.}
\begin{tabular}{lp{1.0 cm}lp{0.5 cm}cp{0.5 cm}cp{0.5 cm}c}
\hline
\hline
Defect & & \multicolumn{7}{c}{$\Delta \Omega$ (eV)} \\
       & & 3C-SiC bulk && C-coated [100] SiC NW  && Si-coated [100] SiC NW  &&  [111] SiC NW  \\
\hline
C$_{\rm Si}$  &&4.22/2.76 && 4.51/3.05 &&5.26/3.80 && 4.05/2.59  \\
C -- C        &&7.75/7.02 &&6.78/6.05 &&4.94/4.21 && 7.09/6.36\\
C -- C$_{\rm Si}$ &&7.97/5.78 &&7.36/5.17 &&5.25/3.06 &&7.52/5.33  \\
C -- Si &&8.35/7.62 &&7.53/6.80 &&4.64/3.91 &&7.45/6.72 \\
(C -- C)$_{\rm i}$  &&9.52/8.06 &&8.05/6.59 &&5.15/3.69 &&8.07/6.61  \\
2(C -- C)$_{\rm C}$ &&12.15/10.69 &&10.67/9.21 &&6.15/4.69 &&10.56/9.10 \\
$[$(C -- C)$_{\rm Si}$ + (C -- C)$_{\rm C}$$]$ &&10.97/8.05 &&9.87/6.95 && 6.48/3.56 &&9.95/7.03  \\
2(C -- C)$_{\rm Si}$ &&13.74/9.36 &&12.03/7.65 && 6.80/2.42 &&12.12/7.74  \\
\hline
\hline
\end{tabular}
\label{tab1}
\end{table*}

In 3C-SiC bulk, and in C-coated [100] and [111] SiC NWs, the C$_{\rm
  Si}$ defect [Fig.~\ref{fig2}(a)] is the energetically most favorable
configuration: we find $\Delta \Omega$=2.76, 3.05, and 2.59~eV, respectively. 
Further formation energy comparison indicates that, upon presence of
interstitial carbon atoms, C -- Si [Fig.~\ref{fig2}(d)] is
energetically more stable than (C -- C)$_{\rm i}$
[Fig.~\ref{fig2}(e)], in 3C-SiC bulk, for any C concentration.  We
find formation energy differences ($\Delta\Omega$), between C -- Si and
(C -- C)$_{\rm i}$, of 1.17, 0.81 and 0.44~eV, at Si-rich,
stoichiometric and C-rich conditions, respectively.  That is, the
formation of (C -- C)$_{\rm i}$ structures is quite unlikely in
3C-SiC.  Nevertheless, in both C-coated [100] and [111] SiC NWs, (C --
C)$_{\rm i}$ is more favorable than C -- Si at C-rich limit (by
0.21~eV in C-coated [100] SiC NW and 0.11~eV in [111] SiC NW).
In this case, in contrast with the 3C SiC bulk phase, we
may find C -- C interstitial dimers embeded in SiC NWs
[Fig.~\ref{fig2}(e)]. However, by reducing the concentration of C atoms, we verify that C --
Si becomes less stable than (C -- C)$_{\rm i}$ for $\Delta \mu_{\rm C}
\geq -0.21$ and $-0.11$~eV, in C-coated [100] and [111] SiC NWs,
respectively. 

\begin{figure}
\includegraphics[scale=0.60]{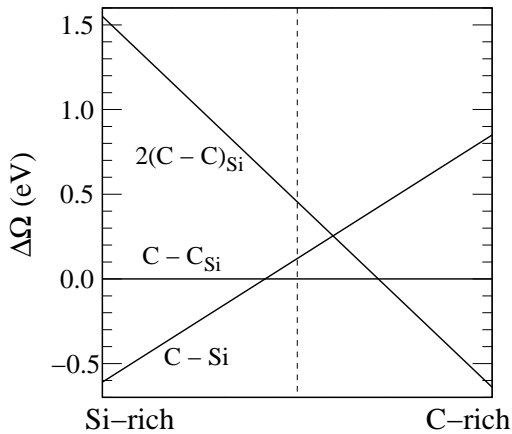}
\caption{\label{fig3}
Relative formation energies ($\Delta \Omega$) of the most stable C defects in Si-coated [100] SiC
NW. The vertical dashed line corresponds to the stoichiometric condition.}
\end{figure}

  Different from C-coated [100] and [111] SiC NWs, where C$_{\rm Si}$
  is the energetically most favorable configuration throughout the
  allowed range for the C chemical potential, in Si-coated [100] SiC
  NW the most stable defect depends on the C concentration (see Table
  \ref{tab1}). At Si-rich limit C -- Si is the most favorable
  configuration, followed by the C -- C and (C -- C)$_{\rm i}$
  defects. However, under C-rich conditions 2(C -- C)$_{\rm Si}$ is
  the most stable defect, followed by the C -- C$_{\rm Si}$ and $[$(C
  -- C)$_{\rm Si}$ $+$ (C -- C)$_{\rm C}$$]$ aggregates. 
  Figure~\ref{fig3} summarizes our calculated formation energy
    results. The energetically most favorable defect in Si-coated
  [100] SiC NW is C -- Si for $\Delta \mu_{\rm C} \leq -0.43$~eV, C --
  C$_{\rm Si}$ for $-0.43 \leq \Delta \mu_{\rm C} \leq -0.21$~eV and
  2(C -- C)$_{\rm Si}$ for $\Delta \mu_{\rm C} \geq -0.21$~eV. At
  stoichiometric conditions, C -- C$_{\rm Si}$ is the most stable
  configuration, followed by C -- Si and (C -- C)$_{\rm i}$. We find
  total energy differences, under stoichiometric conditions, of
  $0.12$~eV between C -- C$_{\rm Si}$ and C -- Si, and of $0.27$~eV
  between C -- C$_{\rm Si}$ and (C -- C)$_{\rm i}$.

\begin{figure}
\includegraphics[scale=0.45]{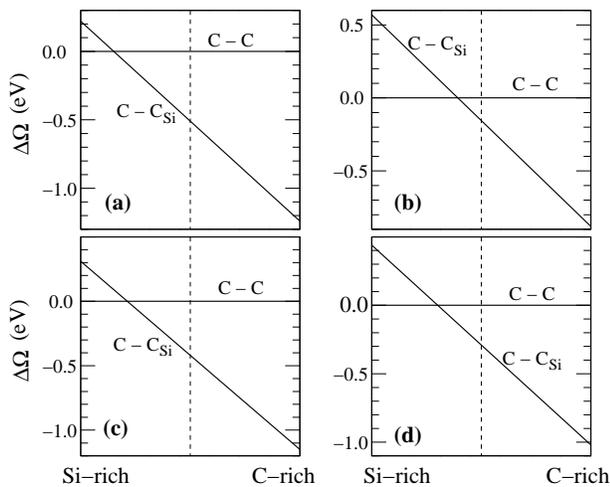}
\caption{\label{fig4}
Relative formation energies ($\Delta \Omega$) of C -- C and C -- C$_{\rm Si}$ defects in (a) 3C-SiC bulk, (b) C-coated [100]
SiC NW, (c) Si-coated SiC NW and (d) [111] SiC NW. The vertical dashed lines indicate the stoichiometric condition.}
\end{figure}

  According to Ref.~\onlinecite{gali}, C -- C$_{\rm Si}$ is
  energetically most favorable than C -- C in 3C-SiC bulk. This result
  can be explained by the larger relaxation required to put two C
  atoms in a C site than to put two C atoms in a Si site.
In the present work we observe a similar behavior [Fig.~\ref{fig4}]:
under C-rich conditions C -- C$_{\rm Si}$ is the most stable complex
with one carbon interstitial in 3C-SiC bulk and in [100] and [111] SiC
NWs.  However, C -- C becomes energetically more stable than C --
C$_{\rm Si}$ for low carbon concentration: $\Delta \mu_{\rm C}\leq
-0.62$~eV (in 3C-SiC bulk), $-0.44$~eV (in C-coated [100] SiC NW),
$-0.57$~eV (in Si-coated SiC NW) and $-0.51$~eV (in [111] SiC NW).  At
stoichiometric conditions, C -- C$_{\rm Si}$ is more stable than C --
C by 0.52~eV (in 3C-SiC bulk), 0.16~eV (in C-coated [100] SiC NW),
0.42~eV (in Si-coated SiC NW) and 0.29~eV (in [111] SiC NW). Our
result for 3C-SiC bulk is in close agreement with the result reported
in Ref.~\onlinecite{gali}, where C -- C$_{\rm Si}$ is more favorable
than C -- C by 0.5~eV, at stoichiometric conditions.

Among the defects with two carbon interstitials, 2(C -- C)$_{\rm Si}$
is not expected to occur in 3C-SiC bulk, C-coated [100] SiC NW and
[111] SiC NW, but is the most stable defect, at C-rich limit, in
Si-coated [100] SiC NW. Besides that, at stoichiometric condition, the
formation energy of 2(C -- C)$_{\rm Si}$ is 11.55, 9.84 and 9.93~eV in
3C-SiC bulk, C-coated [100] SiC NW and [111] SiC NW, respectively, and
4.61~eV in Si-coated [100] SiC NW. One possible reason for that is the
larger relaxation of the Si-coated [100] SiC NW due to its
Si-terminated surface.  Hence, the 2(C -- C)$_{\rm Si}$ complex is
created with small stress in the lattice. This surface effect is
expected to disappear for larger nanowires.

\begin{table*}
\caption{Binding energies (in eV) of C defects in 3C-SiC
bulk, and hydrogen-passivated [100] and [111] SiC NWs. In each column the values correspond to
the stoichiometric condition.
}
\begin{tabular}{lp{0.8 cm}cp{0.4 cm}cp{0.4 cm}cp{0.4 cm}c}
\hline
\hline
Cluster & & \multicolumn{7}{c}{Binding energy (eV)} \\
       & & 3C-SiC bulk && C-coated [100]  && Si-coated [100] &&  [111] SiC NW \\ 
       & &            &&   SiC NW         &&         SiC NW  &&               \\
\hline
C$_{\rm Si}$$+$$[$C -- C$]$ $\rightarrow$ $[$C -- C$_{\rm Si}$$]$  && -4.00  && -3.93 &&
-4.95 && -3.62   \\
C$_{\rm Si}$$+$$[$C -- Si$]$ $\rightarrow$ $[$C -- C$_{\rm Si}$$]$  && -4.60  && -4.68 &&
-4.65 && -3.98   \\
$[$C -- C $]$ $+$ $[$C -- C$]$  $\rightarrow$ $[$(C -- C)$_{\rm i}$$]$ && -5.98 && -5.51
&& -4.73 && -6.11  \\
$[$C -- C $]$ $+$ $[$C -- C$]$  $\rightarrow$ $[$2(C -- C)$_{\rm C}$$]$ && -3.34 && -2.88
&& -3.72 && -3.61  \\
$[$C -- C$_{\rm Si}$$]$ $+$ $[$C -- C$]$ $\rightarrow$ $[$(C -- C)$_{\rm Si}$ + (C -- C)$_{\rm
C}$$]$ && -4.75 && -4.27   && -3.71  && -4.66  \\
$[$C -- C$_{\rm Si}$$]$ $+$ $[$C -- C$_{\rm Si}$$]$ $\rightarrow$ $[$2(C -- C)$_{\rm Si}$$]$  &&
-2.20  && -2.69 && -3.70 && -2.92  \\
2$[$(C -- C)$]$ $+$ 2C$_{\rm Si}$ $\rightarrow$ $[$2(C -- C)$_{\rm Si}$$]$ && -10.20 &&
-10.55 && -13.60 && -10.16 \\
\hline
\hline
\end{tabular}
\label{tab2}
\end{table*}

We next examine the energy  gain upon the formation of C clusters
  through the combination of interstitial  C atoms.  In this case, the
  energy gain  was determined by  comparing the total energies  of the
  separated  systems ($E_i$)  and the  total  of  a given  (stoichiometrically
  equivalent) C cluster ($E_c$). We define the binding energy 
of the C cluster as $E^b = E_c - E_i$.
Our calculated  binding energies are  summarized in Table~\ref{tab2}.
The results presented in  each column correspond to the stoichiometric
condition.  In 3C-SiC  bulk or SiC NW with  a carbon antisite (C$_{\rm
  Si}$ defect),  a carbon interstitial can  be captured by  C site, Si
site or C antisite.  The last process is more favorable than the first
(second) one by 4.00~eV (4.60~eV) in 3C-SiC bulk, 3.93~eV (4.68~eV) in
C-coated [100]  SiC NW, 4.95~eV  (4.65~eV) in Si-coated [100]  SiC NW,
and  3.62~eV  (3.98~eV)  in  [111]  SiC  NW.  Considering  two  carbon
interstitials,  the  formation  of  carbon clusters  is  energetically
favored in  3C-SiC bulk and in both  [100] and [111] SiC  NW. In fact,
the aggregates  (C -- C)$_{\rm  i}$ and 2(C  -- C)$_{\rm C}$  are more
stable  than two  isolated  C --  C  defects by  5.51  and 2.88~eV  in
C-coated  [100] SiC  NW, and  by  6.11 and  3.61~eV in  [111] SiC  NW.
Still, in  Si-coated [100] SiC NW  the energy gain is  13.6~eV to form
2(C -- C)$_{\rm Si}$ from isolated  (C -- C) and C$_{\rm Si}$ defects,
and 3.7~eV  to form 2(C  -- C)$_{\rm Si}$  from two isolated C -- C$_{\rm  Si}$. In
3C-SiC  bulk these  energy gains  are 10.2  and  2.2~eV, respectively.
Here we  can infer that, in general, the  formation of C clusters
  is  quite  likely in  SiC  NWs,  in  particular for  thin  Si-coated
  [100] SiC NWs.  It is worth to  note that, in order to get a complete
  picture of the formation of C clusters, the C diffusion mechanism is
  an important issue, however, it is beyond the scope of the present
  work.

\begin{figure}[!h]
\includegraphics[scale=0.10]{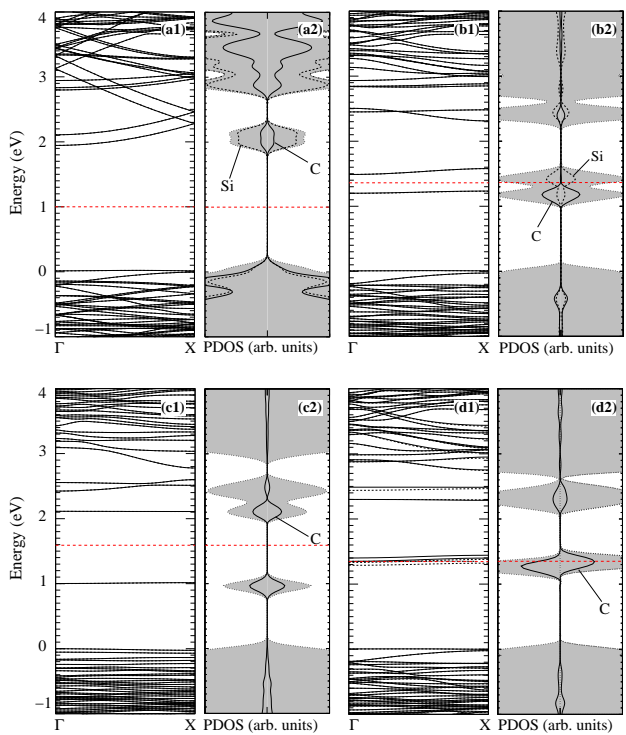}
\caption{\label{pdos-nano100Si}   (Color   online)   Electronic   band
  structure  and the projected  density of  states (PDOS)  of Si-coated
  [100] SiC NW. (a) Pristine NW, and with the presence of defects, (b)
  C - Si,  (c) 2(C - C)$_{\rm  Si}$, and (d) C --  C$_{\rm Si}$. In
  the  PDOS  diagrams, the  shaded  regions  correspond  to the  total
  density of states; in  the electronic band structure diagrams, solid
  lines  (dotted lines) represent  the spin-up  (spin-down) electronic
  channels. The  red dashed  lines correspond to  the position  of the
  Fermi  level and  the zero  energy corresponds  to the  valence band
  maximum.}
\end{figure}

 Focusing  on the electronic  properties, we find that  similar to
  the  3C-SiC bulk  phase,\cite{gali} C$_{\rm  Si}$ is  an  electrically inactive
  defect in SiC NWs.  C$_{\rm Si}$ does not introduce states within
  the  fundamental  band  gap.   On the  other  hand,  the
  formation of C clusters gives rise to electronic states lying within
  the  energy  bandgap.   Figure~\ref{pdos-nano100Si}(a) presents  the
  electronic band structure and the projected density of states (PDOS)
  of the pristine Si-coated [100] SiC NW. The pristine system exhibits
  an  energy bandgap (at  the $\Gamma$  point) of  1.98~eV, calculated
  within  our  DFT-LSDA  approximation,  where  the  highest  occupied
  (lowest  unoccupied) states  are mostly  composed  by C  2p (Si  3p)
  orbitals.   The  formation of  C  clusters  give  rise to  localized
  electronic   states    within   the   bandgap,    as   depicted   in
  Figs.~\ref{pdos-nano100Si}(b)--\ref{pdos-nano100Si}(d).    C  --  Si
  defect gives  rise to an  occupied (empty) state at  1.2~eV (1.5~eV)
  above the valence  band maximum [Fig~\ref{pdos-nano100Si}(b1)].  The
  dispersionless  character  of  those  states,  along  the  $\Gamma$X
  direction  ({\it i.   e.}   parallel to  the  NW growth  direction),
  indicate that  those defects  states are localized  around C  -- Si.
  Indeed,  the PDOS  diagram  [Fig.~\ref{pdos-nano100Si}(b2)] indicate
  that  there is a  significant contribution  from the  interstitial C
  atom to  the occupied  defect state, while  the nearest  neighbor Si
  atom contributes  to the formation  of the lowest  unoccupied defect
  state.  At the C-rich limit, the  2(C -- C)$_{\rm Si}$ defect is the
  most favorable  one.  Its electronic  band structure is  depicted in
  Fig.~\ref{pdos-nano100Si}(c1).   We find an  occupied state  at 1~eV
  above the  VBM, and three  unoccupied states lying within  an energy
  interval of 2.1  and 2.6~eV above the VBM.   Those defect states are
  mostly     localized    around     the    interstitial     C    atoms
  [Fig.~\ref{pdos-nano100Si}(c2)].   Finally, for  C  -- C$_{\rm  Si}$
  (energetically most likely at  the stoichiometric condition) we find
  the formation  of spin-unpaired states  within the nearby  the Fermi
  level,  Figs.~\ref{pdos-nano100Si}(d1) and \ref{pdos-nano100Si}(d2).
  Those  defect states  a localized  around C  interstitial  atoms. In
  summary,  different from C$_{\rm  Si}$ defects,  we verify  that the
  formation of  C clusters in  Si coated [100]  SiC NWs gives  rise to
  deep states  within the  bandgap, which  may acts as  a trap  to the
  electronic carriers.

\section{Conclusions} \label{conclusion}

We performed an {\it ab initio} investigation of small carbon clusters
in SiC bulk  and NWs.  We examined clusters  with carbon interstitials
and  antisites in  3C-SiC bulk  and in  hydrogen-passivated  [100] and
[111] SiC  NWs. We observed that  the composition of  the SiC nanowire
surface strongly  influences the formation of the  carbon clusters. In
fact, C$_{\rm  Si}$ is the energetically  most favorable configuration
in 3C-SiC  bulk, and in C-coated [100]  and [111] SiC NWs,  but is not
expected to  occur in Si-coated  [100] SiC NW. The  energetically most
stable defect in Si-coated [100] SiC NW is C -- Si at Si-rich limit, C
-- C$_{\rm Si}$  at stoichiometric conditions and 2(C  - C)$_{\rm Si}$
under  C-rich   conditions.   

Comparing  the total energies  of the  C --  C and  C --  C$_{\rm Si}$
defects (in 3C-SiC bulk and in  [100] and [111] SiC NWs), we find that
C --  C$_{\rm Si}$ is  more stable than  C -- C at  stoichiometric and
C-rich  limits.   This  finding  is  in  accordance  with  the  larger
relaxation required to put  two C atoms in a C site  than to put two C
atoms  in  a Si  site.   However,  we observed  that  C  -- C  becomes
energetically  more  stable   than  C  --  C$_{\rm  Si}$   for  low  C
concentration. Further total energy calculation indicate that the
  formation of  carbon clusters in 3C-SiC  bulk and in  both [100] and
  [111]  SiC NWs  is energetically  favored.  Finally,  our electronic
  band structure  calculations indicate that  (i)~similar to  the 3C
  SiC  bulk phase,  C$_{\rm  Si}$ defects  are electrically  inactive,
  while (ii)~in Si coated [100]  NWs the C clusters gives rive to deep
  (localized) levels within the NW bandgap.

\section*{ACKNOWLEDGMENTS}

This work received financial support from the Brazilian agencies CNPq and FAPEMIG. 
JMM wishes to thank I. S. Santos de Oliveira for fruitful technical discussions.  
All calculations were performed using the computational facilities of CENAPAD/SP.

\end{document}